\documentstyle[12pt]{article}

\addtocounter{section}{-1}
\catcode `\@=11

\if@twoside
\else
\fi
\def\@maketitle{\newpage \null \vskip 2em   
 \begin{center}
  {\bf \@title \par}     
  \vskip 3em                
  {
    \lineskip .5em \sc          
    \begin{tabular}[t]{c}\@author \end{tabular}\par}
  \vskip 2em              
  {\it \lineskip .5em
    \begin{tabular}[t]{c}\@address \end{tabular}\par}
  \vskip 2em              
  {(Received \@date)}           
 \end{center}
 \par \vskip 2em}                

\def\address#1{\gdef\@address{#1}}

\def\abstract{\if@twocolumn \section*{Abstract}
\else \small
\begin{center}{\bf Abstract\vspace{-.5em}\vspace{0pt}}\end{center}\quotation
\fi}

\def\endabstract{\if@twocolumn\else\endquotation\fi}

\newcounter{figcaption}
\def\thefigcaption{\arabic{figcaption}}
\def\fnum@figcaption{{\bf Fig. \thefigcaption :}}

\def\figcaption{
  \par\pagebreak {\parindent 0pt \bf Figure Captions} \par \vskip 10pt
  \list{\fnum@figcaption}
  {\leftmargin 5em \labelwidth\leftmargin\advance\labelwidth-\labelsep
   \def\makelabel##1{##1\hfil} \usecounter{figcaption}}
}

\def\thereferences#1{\section*{References\@mkboth
 {REFERENCES}{REFERENCES}}\list
 { \arabic{enumi})\ }{\settowidth\labelwidth{#1)\ }\leftmargin\labelwidth
 \advance\leftmargin\labelsep \usecounter{enumi}}
 \def\newblock{\hskip .11em plus .33em minus -.07em}
 \sloppy \sfcode`\.=1000\relax}

\def\jcite{\@ifnextchar [{\@tempswatrue\@jcitex}{\@tempswafalse\@jcitex[]}}

\def\@jcitex[#1]#2{\if@filesw\immediate\write\@auxout{\string\citation{#2}}\fi
  \def\@citea{}\@jcite{\@for\@citeb:=#2\do
    {\@citea\def\@citea{,\penalty\@m}\@ifundefined
       {b@\@citeb}{{\bf ?}\@warning
       {Citation `\@citeb' on page \thepage \space undefined}}%
\hbox{\csname b@\@citeb\endcsname}}}{#1}}

\newfont{\scrptrm}{cmr8}
\def\@jcite#1#2{${}^{\scrptrm {#1\if@tempswa , #2\fi})}$}

\def\acknowledgement{\if@twocolumn \section*{Acknowledgement}
\else \normalsize \begin{center}
{\bf Acknowledgement\vspace{-.5em}\vspace{0pt}} \end{center}\quotation
\fi}

\def\endacknowledgement{\if@twocolumn\else\endquotation\fi}

\def\@normalsize{\@setsize\normalsize{25pt}\xiipt\@xiipt
\abovedisplayskip 12pt plus3pt minus7pt%
\belowdisplayskip \abovedisplayskip
\abovedisplayshortskip  \z@ plus3pt%
\belowdisplayshortskip  6.5pt plus3.5pt minus3pt}

\def\small{\@setsize\small{22.6pt}\xipt\@xipt
\abovedisplayskip 11pt plus3pt minus6pt%
\belowdisplayskip \abovedisplayskip
\abovedisplayshortskip  \z@ plus3pt%
\belowdisplayshortskip  6.5pt plus3.5pt minus3pt
\def\@listi{\parsep 4.5pt plus 2pt minus 1pt
 \itemsep \parsep \topsep 9pt plus 3pt minus 5pt}}

\def\footnotesize{\@setsize\footnotesize{20pt}\xpt\@xpt
\abovedisplayskip 10pt plus2pt minus5pt%
\belowdisplayskip \abovedisplayskip
\abovedisplayshortskip  \z@ plus3pt%
\belowdisplayshortskip  6pt plus3pt minus3pt
\def\@listi{\topsep 6pt plus 2pt minus 2pt\parsep 3pt plus 2pt minus 1pt
\itemsep \parsep}}

\def\scriptsize{\@setsize\scriptsize{15.8pt}\viiipt\@viiipt}
\def\tiny{\@setsize\tiny{11.6pt}\vipt\@vipt}
\def\large{\@setsize\large{30pt}\xivpt\@xivpt}
\def\Large{\@setsize\Large{36.6pt}\xviipt\@xviipt}
\def\LARGE{\@setsize\LARGE{41.6pt}\xxpt\@xxpt}
\def\huge{\@setsize\huge{50pt}\xxvpt\@xxvpt}

\def\section{\@startsection {section}{1}{\z@}{-3.5ex plus -1ex minus
    -.2ex}{2.3ex plus .2ex}{\normalsize\bf}} 
\def\subsection{\@startsection{subsection}{2}{\z@}{-3.25ex plus -1ex minus
   -.2ex}{1.5ex plus .2ex}{\normalsize\it}}  
\def\subsubsection{\@startsection{subsubsection}{3}{\z@}{-3.25ex plus
 -1ex minus -.2ex}{1.5ex plus .2ex}{\normalsize\it}} 

\def\thesection{\S\arabic{section}.}
\def\thesubsection{\arabic{section}.\arabic{subsection}}

\def\appendix{\par
  \@appendix  
  \setcounter{section}{0} \setcounter{subsection}{0}
  \def\thesection{{\bf Appendix} \Alph{section}:}
  \def\thesubsection{\thesection\arabic{subsection}} 
  \@addtoreset{equation}{section}   
  \def\theequation{\Alph{section}.\arabic{equation}} 
}

\def\@appendix{{\addvspace{3ex} \begin{center}{\bf APPENDIX}\end{center}}}

\catcode `\@=12

\def \sech{\mbox{sech }}

\newcounter{tmpequation}

   \def\theequation{\arabic{section}.\arabic{equation}}
\def\tmptheequation{\arabic{section}.\arabic{tmpequation}}

\def\eqnal{
    \addtocounter{equation}{1}
	\setcounter{tmpequation}{\value{equation}}
	\setcounter{equation}{0}

	\let\savetheequation=\theequation
	\renewcommand{\theequation}{\tmptheequation\alph{equation}}
}
\def\endeqnal{
	\setcounter{equation}{\value{tmpequation}}
	\let\theequation=\savetheequation
}

\newenvironment{eqnumalpha}{\eqnal}{\endeqnal}

\newcommand{\bequ}{ \begin{equation} }
\newcommand{\eequ}{ \end{equation} }
\newcommand{\barr}{ \begin{array} }
\newcommand{\earr}{ \end{array} }
\newcommand{\beqarr}{ \begin{eqnarray} }
\newcommand{\eeqarr}{ \end{eqnarray} }

\newcommand{\baralpha}{ \begin{eqnumalpha} \beqarr}
\newcommand{\earalpha}{ \eeqarr \end{eqnumalpha}}

\def\Del0{\Delta_{0}}
\def\vepsi{\varepsilon}
\def\gamm{\gamma}
\def\m1{m_{1}}

\def\Del#1{\Delta_{#1}}
\def\sig#1{\sigma_{#1}}

\def\Gam{\Gamma}
\def\gam1{\gamma_{1}}
\def\al {\alpha}
\def\vf{v_{F}}

\def\omeg0{\omega_{0}}

\def\Omeg0{\Omega_0}

\def\d{{\rm d}}
\def\e{{\rm e}}

\def\i{{\rm i}}

\begin{document}
\title{\large{
{\bf Effects of the Lattice Discreteness on a Soliton }}\\
           \large{ {\bf in the Su-Schrieffer-Heeger Model  } }  }

\author{           Ry\=oen S{\sc hirasaki} \footnotemark[1] }

 \address{   {\sl Department of Physics, Faculty of Education, Yokohama
National University, } \\
 {\sl Yokohama, Kanagawa 240 } }

\date{ August 31,1995 }

\maketitle

{\abstract
 In this paper we analytically study the effects of the lattice discreteness on
the electron band in the SSH model. We propose a modified version of the  TLM
model which is derived from the SSH model using a continuum approximation. When
a soliton is induced in the electron-lattice system, the electron scattering
states both at the bottom of the valence band and the top of the conduction
band are attracted to the soliton. This attractive force induces weakly
localized electronic states at the band edges.  Using the modified version of
the TLM model, we have succeeded in obtaining analytical solutions of the
weakly localized states and the extended states near the bottom of the valence
band and the top of the conduction band. This band structure does not modify
the order parameters.  Our result coincides well with numerical simulation
works.
}
\footnotetext[1]{E-mail : sirasaki@ed.ynu.ac.jp}

\newpage

\section{Introduction}

\label{partI}
The trans-polyacetylene is an ideal material which bears a non-linear
excitation. A soliton is the non-linear excitation which changes the
arrangement of double bonds of carbons.
Su-Schrieffer-Heeger model (SSH model) is the most simple
lattice model which clarifies the soliton.\jcite{latex} In this paper, we study
effects of the lattice discreteness on the electron band and the order
parameter of the soliton in the SSH model.
The Takayama-Lin-Liu-Maki model (TLM model) is known as a continuum version of
the SSH model and gives a analytical solutions for the soliton, the
polaron\jcite{13} and the soliton lattice.\jcite{solito1}
 Takayama {\it et al.} derived this model from the SSH model retaining only the
lowest order term with respect to $a/\xi$, where $a$ is the lattice constant in
the direction of a polymer chain and $\xi$ is the soliton width.
To take into account the discreteness of the SSH model,
 it would be a correct way to include higher order terms which will be treated
in a perturbational way, the TLM model being the unperturbed Hamiltonian.
 Obtained results have to satisfy a selfconsistency condition between the order
parameters and the single-electron wave functions.

There were several works which determined an acoustic component of the order
parameter.
  For example, Maki obtained the order parameters up to the first order of
$a/\xi$ using the TLM model.\jcite{solito2}
At low temperature, soliton friction motion is dominated by the interaction
between the acoustic phonon and the order parameter. Some workers used his
result to study the soliton diffusion.\jcite{solito3}\jcite{solitol}
  Kurita, Ono, and Wada  have taken into account the fact that the band width
is finite,
using a model introduced by Gammel.\jcite{Kurita}\jcite{phase}
It retains the property that the electronic wave number is confined in the
first Brillouin zone.
Thus, the finiteness of the electron band reappears in a natural way even in
the continuum limit.
Kurita {\it et al.} have obtained the first order corrections to the electronic
wave functions and the order parameters by the perturbation method.
 They have found
that there is a divergence in the second order calculations.

It has been pointed out that the first order acoustic component of the order
parameter is working as an attractive potential to give rise to electronic
localized levels at the bottom and the top of the
band.\jcite{Kurita}\jcite{R.S}
  There have been several works which have solved the SSH model numerically.
Kurita {\it et al.} have pointed out that there are two electronic levels which
are localized at the soliton or at the polaron,
in addition to the well-known mid-gap states.\jcite{Kurita}
The structure of the electron band for a soliton is shown in Fig.1. One of them
is below the bottom of the valence band and the other above the top of the
conduction band. Fu, Shuai, Liu, Sun, and Hicks have obtained the two localized
levels for the soliton together with two more levels in the energy
gap.\jcite{num}
Kurita {\it et al}. have analyzed the phase shifts of the electronic wave
functions of the SSH model numerically in ref.\cite{Kurita} and showed that
there are two types of electronic states which are classified according to a
property roughly corresponding to parity.
 There are accordingly two types of the phase shift. They agree with each other
except at band edges.
 The present author and Wada showed in ref.\cite{R.S} that the weakly localized
states at the edges of the band can be determined analytically by solving
equations given by the sum of the TLM Hamiltonian and the first order
correction terms without using the perturbation method.\jcite{R.S.P}

 The purpose of this paper is to develop the method to obtain the scattering
states of electrons near the band edges as well as the localized states
analytically up to the first order of $a/\xi$.
We propose a modified version of the continuum model which is constructed
introducing the electronic wave functions at the edges of the band in addition
to the electron wave functions at the fermi level. The derived equation for the
electron wave function contains the acoustic component of the order parameter
as the attractive potential. The selfconsistency condition is reconstructed
using both the electron wave functions at the edges of the band and at the
fermi level. Solving the equation, we obtained analytical expressions of the
electronic wave functions. Using this model, we will show that the behavior of
the phase shift of the extended electronic states for a soliton and the
existence of the weakly localized states are consistent with the small
distortion of the acoustic component of the order parameter.

 In \S 2, the formulation of the modified version of the TLM model is
constructed.  The self-consistency between the lattice order parameter and the
electron wave function will be shown in \S 3.
 We solve the equations in \S 4 and obtain the structures of electronic bands.
The phase shifts of the extended states are also derived.

\section{Continuation of the Su-Schrieffer-Heeger model}

\label{chap2}

     The Hamiltonian of the SSH model\jcite{latex} is
\beqarr
H_{\rm SSH} & = & \sum_{n,s}t_{n+1,n}(C^{\dagger}_{n+1,s}C_{n,s}
                      +C^{\dagger}_{n,s}C_{n+1,s}) \nonumber \\
& &+{1\over2}K\sum_{n}(u_{n+1}-u_{n})^{2}+{1\over2}M\sum_{n}\dot{u}_{n}^{ 2} ,
\label{we:1}
\eeqarr
where the overlap integral $t_{n+1,n}$ is given by
 $t_{n,n+1}=t_{0}-\alpha(u_{n+1}-u_{n})$.
$t_{n,n+1}$ is
a linear function of the bond length between the $n$ and $n+1$ sites.
The quantity $C^{\dagger}_{n,s}$ is the creation
operator of a $\pi$ electron at site $n$ with spin $s$, and $u_{n}$ is the
displacement of the $n$-th site in the polymer chain. The second term is the
harmonic potential for the bond between the $n$ and $n+1$ sites with the spring
constant $K$ due to the $\sigma$ electrons. The third term is the kinetic
energy of the unit, $M$ being the mass of the unit.

The bond variable $y(n)$ is defined by
\bequ
y(n) = u_{n+1} - u_{n}.
\eequ
 Considering that the dynamics of lattice is sufficiently slow in comparison to
that of the electrons at the Fermi level, we have, in the adiabatic limit,
$$
\dot{u}_{n} = 0 .
$$
Then, the set of electronic wave functions $\Phi_{k,s}(n)$ is determined as
eigen-functions of the equations
\begin{equation}
E_{k}\Phi_{k,s}(n)= -\{t_{0} - \alpha y(n-1) \} \Phi_{k,s}(n-1) - \{t_{0} -
\alpha y(n) \} \Phi_{k,s}(n+1),
\label{we:2}
\end{equation}
with energy $E_{k}$. This equation is derived by the variational principle with
respect to $C_n^\dagger$.
Using $\Phi_{k,s}$, the total energy  is given by $\sum_{k,s} '
\langle\Phi_{k,s} |H_{SSH} | \Phi_{k,s} \rangle $. This is minimized when
$y(n)$
satisfies
\begin{equation}
y(n)= - { \alpha \over K} {\sum_{k,s}}^{'}\{ {\Phi_{k,s}
}^*(n)\Phi_{k,s}(n+1)+c.c \}+{\alpha \over NK} \sum_{n} {\sum_{k,s}}^{'}\{
{\Phi_{k,s}}^*(n)\Phi_{k,s}(n+1)+c.c \},
\label{ye:1}
\end{equation}
where $\sum^{'}$ means a summation over the occupied states. The second
term comes from the constraint
\begin{equation}
\sum_{n}y(n)=0.
\label{ye:2}
\end{equation}
 As is shown by Su-Schrieffer-Heeger,\jcite{latex} this model bears the bond
order wave (BOW) state, which has the bond alternation of the wave number
${\pi}/{a}$. In this state, the bond variable $y(n)$ is given by
\bequ
y_n=\frac{1}{2\alpha}(-)^n \tilde{\Delta}
\label{o1}
\eequ
 where $\tilde{\Delta} $  is the optical component of the order parameter. In
the SSH model, the quantity $\tilde{\Delta}$ is determined by the
selfconsistency equation,
\bequ
1=2\lambda \int_0^{k_F a} \frac{2t_0 \cos ^2 ka}
{
\sqrt{(2t_0)^2\cos^2 ka + \tilde{\Delta}^2 \sin ^2 ka }
} d(k a) .
\label{se2}
\eequ

 When the non-linear excitation is induced in the BOW state, the order
parameter $\tilde{\Delta}$ is distorted near the center of the excitations.
Maki\jcite{solito2}
 had pointed out that the acoustic component of the order parameter also
suffers a distortion around the non-linear
excitation.
The bond variable is written as follows:
\bequ
y_{n} = \frac{1}{2\alpha} \biggl\{
\tilde{\zeta}(na+\frac{a}{2})+(-)^{n}\tilde{\Delta}(na+\frac{a}{2}
)\biggr\},
\label{y1}
\eequ
 where $\tilde{\zeta}$ is the acoustic component of the order parameter.
 Takayama-Lin-Liu-Maki have written the electron wave function nearly at the
Fermi level by the summation over the left-going and the right-going waves,
\jcite{solito1}
\bequ
\Phi_{k,s}(na)=[W_{s}(na) \exp \{\i k_{F}na \}-\i V_{s}(na) \exp \{-\i k_{F} na
\}]\sqrt{a},
\label{ph:1}
\eequ
with the Fermi wave number $k_{F}=\pi/2a$. On the other hand, near the bottom
or the top of the electron band, the electron wave function should be given by
\beqarr
\Phi_{k,s}(na) & = & \{Z_s(na) -\i Y_s(na) \e^{\i 2\ k_F (na) }\}\sqrt{a} .
\label{ph:2} \\
  & & \mbox{for } \ |E| \sim 2t_0 , \nonumber
\eeqarr
where $Z$ and $Y$ are slowly varying functions. Using $W$, $V$, $Z$ and $Y$,
the electronic states in the presence of the non-linear excitation will be
analyzed.
Substituting eqs.(\ref{y1}) and (\ref{ph:1}) into eq.(\ref{we:2}), and
introducing $\psi_{k,s}(x)$ by
\bequ
\psi_{k,s}(x)=\pmatrix{ W_{s}(x) \cr
		        V_{s}(x) \cr
                      } ,
%
%
\label{ps:4}
\eequ
we obtain the equation for the electron,
\beqarr
E_k\psi _{k,s} (x) &  = & -\i \Delta_0 \sigma_3 \partial \psi_{k,s} (x)+
\tilde{\Delta}
(x) \sigma_1 \psi_{k,s} (x)
\nonumber \\
 & & + m_1 \biggl\{ \frac{1}{2}(\partial\tilde{\zeta}(x)) \sigma_3 \psi_{k,s} +
\i \tilde{\zeta}(x) \sigma_3 (\partial \psi_{k,s} (x))
\biggr\}
 +\dots ,
\label{q1}
\eeqarr
 where $x=na $, and $2\Del0$ being the electron band gap.  The quantity
$m_1=a/\xi=\Del0/(2t_0)$ is the smallness parameter.
The differentiation $\partial$ is defined by $\partial =\xi \frac{\partial
}{\partial x}$ where $\xi=2t_0a/\Del0 $ is soliton width.  In the calculation,
$\psi_{k,s}(x\pm a)$,  the functions $\tilde{\Delta}(x\pm a/2)$ and
$\tilde{\zeta}(x\pm a/2)$ were expanded around $x$.

Substituting eqs.(\ref{y1}) and (\ref{ph:2}) into eq.(\ref{we:2}) and
 introducing $\varphi_{q,s}(x)$ by
\bequ
\varphi_{q,s}(x)=\pmatrix{ Z_{s}(x) \cr
		       Y_{s}(x) \cr
                      } ,
\label{ps:5}
\eequ
 we obtain the equation for the electron near the edges of electron band,
\beqarr
E_q \varphi_{q,s} (x) &  = & \biggl(-2t_0 -\frac{\Del0}{2}m_1
\partial^2 +\tilde{\zeta}(x)\biggr) \sig3\varphi_{q,s} (x)
\nonumber \\
 & & -\i m_1 \biggl(\frac{1}{2}(\partial\tilde{\Delta}(x) )
+\tilde{\Delta} \partial \biggr)\sig1 \varphi_{q,s} (x)
+\dots .
\label{q2}
\eeqarr
 The functions $\varphi_{q,s}(x\pm a)$,  $\tilde{\Delta}(x\pm a/2)$ and
$\tilde{\zeta}(x\pm a/2)$ were expanded around $x$.
  Equations(\ref{ps:4}), (\ref{q1}), (\ref{ps:5}) and (\ref{q2}) are the
equations of the electron wave function.

Substituting eqs.(\ref{ph:1}) and (\ref{ph:2}) into eq.(\ref{ye:1}), we obtain
the selfconsistency relation between the electron and the bond variable $y(n)$.
Using eq.(\ref{y1}) and  equating the slowly varying term and fast oscillating
term by $(-)^n$, we obtain the equations for the order parameters
 $\tilde{\Delta}(x)$ and $\tilde{\zeta}(x)$
\beqarr
\tilde{\Delta}(x) & = & -\frac{4\alpha^2 a}{K} {\sum_{k,s} }'
 \psi_{k,s}^\dagger (x) \sig1 \psi_{k,s} (x)
\nonumber \\
 & & -m_1 \frac{2\alpha^2 a}{K} {\sum_{q,s} }'
\i \{ (\partial \varphi_{q,s}^\dagger (x) )  \sig1 \varphi_{q,s}
(x)-\varphi_{q,s}^\dagger (x) \sig1
 ( \partial\varphi_{q,s}(x) ) \} ,
\label{q3}
\eeqarr
\beqarr
\tilde{\zeta}(x) & = & -\frac{4\alpha^2 a}{K} {\sum_{q,s} }'
 \varphi_{q,s}^\dagger (x) \sig3 \varphi_{q,s} (x)
\nonumber \\
 & & + m_1 \frac{2\alpha^2 a}{K} {\sum_{k,s} }'
\i  \{
(\partial \psi_{k,s}^\dagger (x) )  \sig3\psi_{k,s} (x)-
\psi_{k,s}^\dagger (x) \sig3
 ( \partial\psi_{k,s} (x) ) \} ,
\label{q4}
\eeqarr
where the summation ${\sum}'$ is performed over the occupied states.
$\varphi_{q,s}(x\pm a/2)$ and $\psi_{k,s}(x\pm a/2)$ were expanded around $x$.

Equation (\ref{q1}) determines the wave function of the electron at the Fermi
level, while
eq.(\ref{q2}) is the wave function of the electron near the top of the
conduction or the bottom of the valence band.
  Equations (\ref{q3}) and (\ref{q4}) are the selfconsistency conditions for
the order parameters $\tilde{\Delta}$ and $\tilde{\zeta}$.

  The zeroth order terms of the expansion with respect to $m_1$ in
eqs.(\ref{q1}) and (\ref{q3}) give the TLM model.
 Equations (\ref{q2}), (\ref{q4}) and the higher order terms with respect to
$m_1$ in eqs.(\ref{q1}) and  (\ref{q3}) come from the lattice discreteness of
the SSH model.

The order parameter of soliton in the TLM model is given by \jcite{solito1}
\bequ
\tilde{\Delta}(x)  = \Del0 \tanh \kappa x ,
\label{so1'}
\eequ
with $\kappa=1/\xi$.
 The electron wave function for the scattering state is \jcite{solito1}
\beqarr
\psi_{\gamma}(x)&=&\frac{N_{\gamma} }{2} \left( \matrix{
                       E_\gamma+\gamma+\i\tanh(\kappa x)\cr
                    \i (E_\gamma-\gamma-\i\tanh(\kappa x) ) \cr
           	                                     } \right) \e^{\i\gamma x}
\nonumber \\
           & & \nonumber \\
 N_\gamma & = & (L(1+\gamma^2)-1/\kappa)^{(-1/2)},
\nonumber  \\
 E_\gamma  & = & \pm \sqrt{1+\gamma^2},
 \label{w1}
\eeqarr
\bequ
\gamma=\xi k,
\label{w1s}
\eequ
 where $k$ is the wave number. In eq.(\ref{w1}), the  plus sign corresponds to
the conduction band and the minus sign to the valence band.
The energy of the electron is given by
\bequ
E_k= \Del0 E_\gamma .
\label{Ek}
\eequ
 The electronic localized state inner side of the band gap is given by
\beqarr
\psi_{B0} (x)&=&\frac{ 1 }{2\sqrt{ \xi } } \sech\kappa x
    \left( \matrix{
                     1   \cr
                    -\i\cr
           	                                     } \right) ,
\label{w2} \\
           & & \nonumber \\
  E_M  & = & 0 .
 \eeqarr
  The quantity $\Del0$ is determined by the lowest order term of the
selfconsistency equation (\ref{q3}). This condition becomes
\bequ
1=2\lambda \log\Biggl( \frac{2\times 2t_0}{\Del0}
\Biggr) Ê,
\label{se1}
\eequ
with $\lambda=4\alpha^2/(2t_0K \pi)$.

Maki  had analyzed the soliton using eqs.(\ref{q1}), (\ref{q3}) and
(\ref{q4}),\jcite{solito2} neglecting the terms related to $\varphi_{q,s}$ and
retaining terms up to the first order with respect to $\m1$. He obtained the
acoustic order parameter
 \bequ
\tilde{\zeta}(x)  =  -\frac{\m1}{2} \Del0 \sech^2 \kappa x ,
+O({m_1}^2) .
\label{so1}
\eequ
The optical order parameter and the electronic wave function were unchanged.
 Numerical and analytical works in the SSH model have pointed out that there
are weakly localized states Êaround the soliton in addition to the mid-gap
state. Their energy levels are located at the top of the conduction band and
the bottom of the valence band.\jcite{Kurita}\jcite{R.S}\jcite{num}
 The acoustic component of the order parameter $\tilde{\zeta}$ works as an
attractive potential at the soliton center, making one  weakly localized state
at each of the band edges.\jcite{R.S}
 As shown in Fig.1, the level structure is different from that obtained in
ref.\cite{solito2}.
Therefore, it is important to see whether the selfconsistency between
eq.(\ref{so1})  and the electronic states is satisfied or not.
In the next section, we obtain the electron wave function near the edges of the
electron bands and the selfconsistency will be confirmed up to the first order
with respect to $\m1$.

\section{Selfconsistency}
\setcounter{equation}{0}

 In the present section, the electronic states and the lattice configuration
are self consistently determined, using eqs.(\ref{q1})  $ \sim $ (\ref{q4}).
We expand the electron wave function by the smallness parameter $\m1$,
\beqarr
 \psi_{k,s}(x) & = & \psi_{k,s}^{(0)} (x) + \m1 \psi_{k,s}^{(1)} (x)+ \dots ,
 \nonumber \\
 \varphi_{q,s}(x) & = & \varphi_{q,s}^{(0)} (x)  + \m1 \varphi_{q,s}^{(1)} (x)
+\cdots,
\label{r0}
\eeqarr
and the order parameters, in the same way,
\beqarr
\tilde{\Delta}(x) & = & \Delta^{(0)} (x) + \m1 \Delta^{(1)} (x) +\cdots,
\nonumber \\
\tilde{\zeta} (x) & = & \zeta^{(0)} (x) +\m1\zeta^{(1) } (x) +\cdots.
\label{r0'}
\eeqarr
Then, the equations for electrons, eqs.(\ref{q1}) and (\ref{q2}), become
\beqarr
E_k^{(0)}\psi_{k,s}^{(0)} & = & H^{(0)} \psi_{k,s}^{(0)},
\label{r00} \\
E_k^{(1)}\psi_{k,s}^{(0)}+E_k^{(0)}\psi_{k,s}^{(1)}  & = & H^{(1)}
\psi_{k,s}^{(0)}
+H^{(0)} \psi_{k,s}^{(1)},
\label{r01} \\
E_q^{(0)} \varphi _{q,s}^{(0)} & = & h^{(0)} \varphi _{q,s}^{(0)},
\label{r02} \\
E_q^{(1)} \varphi _{q,s}^{(0)}+E_q^{(0)} \varphi _{q,s}^{(1)}  & = & h^{(1)}
\varphi _{q,s}^{(0)}
+h^{(0)} \varphi _{q,s}^{(1)},
\label{r03}
\eeqarr
 where the Hamiltonian $H^{(i)}$, $h^{(j)}$ $(i,j=0,1)$ are given as follows:
\beqarr
H^{(0)} & = & -\Del0 \i \sig3\partial + \Delta^{(0)}(x) \sig1 ,
\label{r1} \\
H^{(1)} & = & \i \biggl\{ \frac{1}{2} (\partial \zeta^{(0)}(x))
 + \zeta^{(0)}(x) \partial \biggr\} \sig3 ,
\label{r2} \\
h^{(0)} & = &  \biggl\{ -2t_0 -\frac{\m1}{2} \Del0 \partial^2+
\zeta^{(0)} (x) +\m1 \zeta^{(1)} (x) \biggr\}
 \sig3 ,
\label{r4} \\
h^{(1)} & = & -\i \biggl\{ \frac{1}{2} (\partial \Delta^{(0)} (x))
 + \Delta^{(0)} (x) \partial \biggr\} \sig1 .
\label{r5}
\eeqarr
Because the electrons near the edges of the electron bands suffer the
attractive potential given by the acoustic component of the order parameter
$\tilde{\zeta}(x)$, the second and the fourth term in the right hand side of
eq.(\ref{r4}) is included in the unperturbed Hamiltonian $h^{(0)}$.
Substituting eqs.(\ref{r0}) and (\ref{r0'}) into eqs.(\ref{q3}) and (\ref{q4}),
the selfconsistency equations become
\beqarr
\Delta^{(0)}(x) & = & -\frac{4\alpha^2 }{K} a {\sum_{k,s} }'
 {\psi_{k,s}^{(0)} } ^\dagger (x) \sig1 \psi_{k,s} ^{(0)} (x),
\label{s1} \\
\Delta^{(1)} (x) & = & -\frac{4\alpha^2 }{K} a {\sum_{k,s} }'
 ({\psi_{k,s}^{(0)} } ^\dagger (x) \sig1 \psi_{k,s} ^{(1)} (x)
+{\psi_{k,s}^{(1)} } ^\dagger (x) \sig1 \psi_{k,s} ^{(0)} (x) )
 \nonumber \\
 & &  -\frac{2\alpha^2 }{K} a {\sum_{q,s} }'\i \{
(\partial { \varphi _{q,s}^{(0)} } ^\dagger (x) )\sig1 \varphi _{q,s} ^{(0)}(x)
- { \varphi _{q,s}^{(0)} } ^\dagger (x) \sig1 (\partial\varphi _{q,s} ^{(0)})
\} ,
\label{s2} \\
 \zeta^{(0)} (x) & = & -\frac{4\alpha^2 }{K} a {\sum_{q,s} }'
 {\varphi _{q,s}^{(0)} } ^\dagger (x) \sig3 \varphi _{q,s} ^{(0)} (x),
\label{s3} \\
\zeta^{(1)} (x) & = & -\frac{4\alpha^2 }{K} a {\sum_{q,s} }'
 \{ {\varphi _{q,s}^{(1)} } ^\dagger (x) \sig3 \varphi _{q,s} ^{(0)} (x)
 + {\varphi _{q,s}^{(0)} } ^\dagger (x) \sig3 \varphi _{q,s} ^{(1)} (x) \}
\nonumber \\
 & & +\frac{2\alpha^2 }{K} a {\sum_{k,s} }'
 \i \{ (\partial {\psi _{k,s}^{(0)} } ^\dagger) (x) \sig3 \psi _{k,s} ^{(0)}
(x)
 - {\psi _{k,s}^{(0)} } ^\dagger (x) \sig3 (\partial \psi _{k,s} ^{(0)} (x)) \}
{}.
\label{s4}
\eeqarr
 Using eqs.(\ref{r00}), (\ref{r02}), (\ref{r1}), (\ref{r4}), (\ref{s1}) and
 (\ref{s3}), an unperturbed solution is obtained.

Equations (\ref{r1}) and (\ref{s1}) are the same equations as in the TLM model.
Thus, in the soliton case, the unperturbed solution $\psi ^{(0)}(x) $  for the
scattering state is given by eq.(\ref{w1}) and
the solution for the mid-gap state is given by eq.(\ref{w2}).
Our model does not modify the electron wave function near the Fermi level which
is given by the TLM model.

The optical component of the order parameter is given by
\bequ
\Delta^{(0)}(x) = \Del0 \tanh \kappa x .
\label{3.14'}
\eequ
The band gap $\Del0$ is given by the equation
 \bequ
 1=2\lambda \log \Biggl( \frac{2\vf\Lambda}{\Del0} \Biggr),
 \label{sc2}
 \eequ
 where $\Lambda$ is introduced as the cut off of the wave number $k$.

The wave function of the electron near the bottom of the valence band or the
top of the conduction band is obtained by eqs.(\ref{r02}) and (\ref{r4}).
The wave function $\varphi$ is given by
\beqarr
\varphi_q^{(0)} (x) &=&
    \left( \matrix{
                     \phi_q (x)   \cr
                   0\cr                          } \right) ,
\label{w4} \\
           & & \nonumber \\
  E_q  & = & -2t_0+\varepsilon_q ,
 \eeqarr
for the valence band, and by
\beqarr
\varphi_q^{(0)} (x) &=&
    \left( \matrix{
                     0 \cr
                   \phi_q (x)   \cr                          } \right) ,
\label{w5} \\
           & & \nonumber \\
  E_q  & = & 2t_0-\varepsilon_q ,
\label{ew5}
\eeqarr
for the conduction band.
 The functions $\phi_q (x) $ are solutions of the Schr\"{o}dinger equation
\bequ
\varepsilon_q \phi_q(x) = \biggl( -\frac{\m1}{2} \Del0 \partial ^2
+\zeta^{(0)}(x)+\m1\zeta^{(1)} (x) \biggr)\phi_q (x) ,
\label{w3}
\eequ
where $\zeta^{(0)}(x)$ and $\zeta^{(1)}(x)$ is the real function given by the
selfconsistency equation (\ref{s3}) and (\ref{s4}).  The quantity
$\varepsilon_q$ is the eigen-value of eq.(\ref{w3}).
As it will be shown later, the forms of $\zeta^{(0)} (x)$ and $\zeta^{(1)} (x)$
do not depend on the structure of $\phi_q (x)$.
Because $\phi_q (x)$ is the solution of the second order differential equation,
there is a completeness relation,
\bequ
\sum_q \phi_q^* (x) \phi_q (y)
=\delta(x-y) .
\label{c1}
\eequ
where $\phi_q (x)$ is renormalized as $\int \phi_q^* (x) \phi_q (x)\d x=1$.  In
eq.(\ref{c1}), the summation is taken over all eigen-functions $\{\phi_q(x)\}$.
{}From the selfconsistency equation (\ref{s3}) and the completeness relation
(\ref{c1}), the zeroth order part of the acoustic component of the order
parameter, $\zeta^{(0)} (x)$, is the constant.
Since, from eq.(\ref{ye:2}),  the integration over $x$ of $\tilde{\zeta}(x)$ is
zero, we obtain
\bequ
\zeta^{(0)}(x) = 0 .
\label{c2}
\eequ
  Therefore, substituting eq.(\ref{c2}) into (\ref{r2}), the first order
Hamiltonian $H^{(1)}$ is zero.  From eq.(\ref{r01}), the first order wave
function is zero.
 \bequ
\psi^{(1)} (x) =0 .
\label{p1'}
\eequ
Substituting eqs.(\ref{w5}) and (\ref{p1'}) into eq.(\ref{s2}),
 the first order term with respect to $\m1$ of the optical component of the
order parameter, $\Delta^{(1)} (x)$, is given by
\bequ
\Delta^{(1)} (x) =0.
\label{d1}
\eequ

{}From eqs.(\ref{r03}) and (\ref{r5}), the first order electron wave function
$\varphi_q^{(1)} (x)$ is given by the perturbation.
\bequ
 \varphi_q^{(1)} (x) = -\i \int_{-\infty}^\infty \d y
Q_q (x,y) \biggl( \frac{1}{2} \sech^2 \kappa x +\tanh\kappa x \partial\biggr)
\sig1 \varphi_q^{(0)} (y) ,
\label{p0}
\eequ
where we used eq.(\ref{3.14'}). $Q_q(x,y)$ is the perturbation kernel,
\bequ
 Q_q (x,y) = \sum_{q'} " \frac{\Del0 \varphi_{q'}^{(0)} (x)
  { \varphi_{q'} ^{(0)} } ^\dagger (y) }{E_q-E_{q'} } .
\label{p1}
\eequ
In the summation ${\sum}"$ over $k'$,  the wave number $q=q'$ where
$sgn(E_q)=sgn(E_{q'})$ is excluded.
Substituting eqs.(\ref{w4}) and (\ref{w5}) into eq.(\ref{p1}), the kernel
$Q_q(x,y)$ for the valence band becomes
\beqarr
Q_q (x,y) & = &
 \left(\begin{array}{cc}
 1 & 0 \\
 0 & 0
\end{array}\right)\Del0
 {\sum_{ k', E_{q'}<0
}
 }" \frac{\phi_{q'} (x) \phi_{q'} ^* (y) }{\varepsilon_q-\varepsilon_{q'} }
-
 \left(\begin{array}{cc}
 0 & 0 \\
 0 & 1
\end{array}\right)
\Del0
 {\sum_{
 k',
 E_{q'}>0
 }
}" \frac{\phi_{q'} (x) \phi_{q'} ^* (y) }{|E_q|
 + | E_{q'} | }
 \nonumber \\
 & \simeq &
 \left( \begin{array}{cc}
 1 & 0 \\
 0 & 0
\end{array}\right)\Del0
 {
\sum_{ k' ,E_{q'}<0}
 }"
\frac{\phi_{q'} (x) \phi_{q'} ^* (y) }{\varepsilon_q-\varepsilon_{q'} }
-
 \left(\begin{array}{cc}
 0 & 0 \\
 0 & 1
\end{array}\right)
\frac{\m1}{2} \delta(x-y) ,
\label{Q1}
\eeqarr
where $\varepsilon_q$ is the eigenvalue of the Schr{\"o}dinger equation
(\ref{w3}). In the calculation from the first line to the second, we
approximated $| E_q | + | E_{q'} |$ by $2 \times (2t_0)$,   since
$\varepsilon_q/(2t_0)$ is the quantity of the order of $\m1 ^2$.  We used the
completeness of  $\phi_k(x)$, eq.(\ref{c1}). As in the same way as
eq.(\ref{Q1}), substitution of eq.(\ref{w4}) into eq.(\ref{p1}) gives the
kernel $Q(x,y)$ for the conduction band,
\beqarr
Q_q(x,y)  & \simeq &
 \left(\begin{array}{cc}
 1 & 0 \\
 0 & 0
\end{array}\right)
 \frac{\m1}{2} \delta(x-y)
-
 \left(\begin{array}{cc}
 0 & 0 \\
 0 & 1
\end{array}\right)
\Del0
{
\sum_{k', E_{q'} > 0 }
}"
  \frac{\phi_{q'} (x) \phi_{q'} ^* (y) }{\varepsilon_q-\varepsilon_{q'} }
 .   \nonumber \\
 & &    \label{Q2}
\eeqarr
{}From eqs.(\ref{p0}), (\ref{Q1}) and (\ref{Q2}), the first order wave function
is
\bequ
\varphi_q ^{(1)} (x) = -\i \ {\rm sgn} (E_q) \frac{\m1}{2}
\biggl( \frac{1}{2} \sech^2 \kappa x +\tanh\kappa x
 \partial \biggr) \sig1 \varphi_q ^{(0)} (x) .
\label{w6}
\eequ
The perturbation Hamiltonian $h^{(1)}$ is expressed by using $\sigma_1$. As we
can see from eqs.(3.17) $\sim$ (3.20), this matrix causes the mixture of the
electron wave functions at the valence band and the wave functions at the
conduction band. Because the energy difference between eqs.(3.18) and (3.20) is
the order of $4t_0$, the perturbation term $\varphi_q^{(1)}(x)$ is the order of
$\Delta_0/(4t_0)\times \varphi_k^{(0)}(x) \sim\m1\varphi_q^{(0)}(x) /2$. Then,
the perturbation Hamiltonian $h^{(1)}$ gives the second order contribution to
the electron wave function $\varphi_q (x)$.
 Substituting eqs.(\ref{w2}), (\ref{w4}) and (\ref{w6}) into eq.(\ref{s4}),
$\zeta^{(1)}(x)$  is given by
\bequ
\zeta^{(1)} (x) = -\frac{\Del0}{2}\sech^2 \kappa x ,
\label{zeta1}
\eequ
while $\varphi_q^{(1)}(x)$ gives no contribution to the acoustic component of
the order parameter $\tilde{\zeta} (x)$.
This result is the same as eq.(\ref{so1}). Because of the completeness of the
electron wave function $\varphi_q(x)$, the electronic states near the edges of
the band do not modify the order parameter up to the first order with respect
to $\m1=a/\xi$.  The order parameters are determined by the electrons near the
Fermi level.

\section{Electronic States at the Band Edges}
\setcounter{equation}{0}
Ê

The electron wave function near the edges of the electron band is given by
eq.(\ref{w4}), where the function $\phi_q (x)$ is the solution of the
Schr\"{o}dinger equation (\ref{w3}). Because the acoustic component of the
order parameter, in the soliton case, is given by eq.(\ref{zeta1}), this
equation is written as follows :
\bequ
\frac{2\varepsilon}{\m1 \Del0} \phi (x) =
\biggl( -\frac{\d ^2 }{\d (\kappa x)^2}-\sech^2 \kappa x \biggr) \phi (x) .
\label{4.1}
\eequ
Using the hypergeometric function, $\phi (x)$ for a positive $\varepsilon$ is
written by
\beqarr
\phi_q (x) & = & \e^{\i q x} F[1+\beta_0, -\beta_0,1+\i \xi q ;
\frac{1+\tanh(\kappa x)}{2}] ,
\nonumber \\
 \frac{2\varepsilon}{\m1 \Del0} & = & (\xi q)^2 .
\label{w7}
\eeqarr
where $q$ is the wave number. $\beta_0$ is defined by eq.(\ref{abd}). The
details of the calculation are given in the Appendix A.
{}From eqs.(\ref{w4}) and (\ref{w7}), the electron energy $E_q$ is given by
\bequ
E_q= \pm 2t_0 \biggl(1-\frac{\m1^2(\xi q)^2}{2} \biggr) ,
\label{e2}
\eequ
As eq.(\ref{4.1}) is symmetric under the transformation $x \rightarrow -x$,
$\phi_q (-x)$ is also a solution with $(2\varepsilon)/\m1\Del0=(\xi q)^2$.
We use combination of $\phi_q (x)$ and $\phi_q (-x)$, which are defined by
\beqarr
\phi_{q,i} (x) & = & \frac{1}{\sqrt{L(|1\pm b(\xi q)| ^2 +| a(\xi q) | ^2 )}}
[\phi_q (x) \pm \phi_q (-x) ] ,
\label{w71}
\eeqarr
where the upper sign corresponds to $i=e$ (even) and the lower sign to $i=o$
(odd).
The normalization factor is calculated in the Appendix B.
A negative $\varepsilon$ corresponds to the localized solution.
It is given by
\bequ
\phi_\ell (x) = \sqrt{\frac{\Gamma(1+2\beta_0)}{\xi \Gamma(1+\beta_0) \Gamma
(\beta_0)} }
\Biggl( \frac{\sech (\kappa x) }{2}\Biggr)^{\beta_0} .
\label{w8}
\eequ
{}From eqs.(\ref{w4}) and (\ref{w8}), the energies of the weakly localized
states are
\bequ
E_\ell = \pm 2t_0 \biggl(  1+\frac{\m1 ^2}{2} \beta_0^2
\biggr).
\label{eell}
\eequ
This form is the same as the eq.(4.28) given in ref.\cite{R.S}.
 In the finite band version of the TLM model, the quantity $\beta_0$ is given
by \jcite{Kurita}
\bequ
\beta_0 \cong \frac{ \sqrt{5-7\lambda}-1 }{ 2 } .
\label{fbbeta}
\eequ
In the present theory, the electron cosine band is linearized. Then,  the
contribution of the order $\lambda$ is dropped from $\beta_0$ and the quantity
$\m1$ in the present theory is a little larger than that obtained
 in the finite band version of the TLM model.

Substituting eqs.(\ref{w71}), (\ref{w8}), (\ref{w4}) and (\ref{w5}) into
eq.(\ref{ph:2}),
 we obtain the electron wave function $\Phi_q$,
\beqarr
\Phi_{q,i} (x) = \sqrt{a} \phi_{q,i} (x) \times \left\{
 \matrix{
               1    & \mbox{for } & E_q <0 ,\cr
               \e^{2\i k_F x }  & \mbox{for } & E_q >0 , \cr
           	                                     } \right.
 \mbox{for the scattering states} & & \nonumber \\
 & & \nonumber \\
\Phi_{\ell} (x) = \sqrt{a} \phi_{\ell} (x) \times \left\{
 \matrix{
               1,    & \mbox{for } & E_\ell <0 ,\cr
               \e^{2 \i k_F x},  & \mbox{for } & E_\ell >0 , \cr
           	                                     } \right.
 \mbox{for the localized states} & & \label{w9}
\eeqarr
with $i=e,o$. The wave function $\Phi_\ell (x)$ of the electron at the bottom
of the valence band is shown in Fig.2.
The band structure is schematically shown in Fig.3. We have selected the cut
off of the wave number $q$ as $\Lambda_\phi={k_F}/{2}={\pi}/(4a)$ and the cut
off of $k$ as $\Lambda=k_F/2$.

\section{Phase Shift of the Extended States in the Electron Band}
\setcounter{equation}{0}

The localized states corresponding to to those discussed in the previous
section were also obtained by the numerical calculation of the soliton in the
SSH model.
 It was shown that there were two branches in the phase shift of the electron
wave function in the SSH model.\jcite{Kurita}
 The phase shift for the electron wave function with even parity approached
$-\pi$ as $q \rightarrow 0$, while that for the  wave function with odd parity
approached $0$ as $q \rightarrow 0$.

 The behavior of the phase shift is reproduced by our analytic wave function,
eq.(\ref{w71}).
  As $|x|\rightarrow \infty$, the electron wave function, eq.(\ref{w71}),
behaves like
\beqarr
\Phi_{q,i} (x) & = & \sqrt{
 \frac{ a }{ L (| 1\pm b(\xi q) |+| a(\xi q) |) }  }
D_n ({\rm sgn}(x) )^i
\nonumber \\
 & & \times [a(\xi q) \e^{\i q|x| }+(b(\xi q) \pm 1 ) \e^{-\i q |x|}] ,
\label{w9'}
\eeqarr
where $i=e, \ o $ denotes even or odd parity. $D_n$ is defined as
\bequ
D_n = \left\{
\begin{array}{lr}
                         1 ,                 & \mbox{for the valence band }, \\
                         \e^{2\i k_F x} , & \ \ \mbox{for the conduction band.
}
\end{array}
\right.
\nonumber
\eequ
 Equation (\ref{w9'}) is rewritten as
\beqarr
\Phi_{q,e} (x) & = & \sqrt{ \frac{a}{ L(| 1+b(\xi q) |^2 +| a(\xi q ) |^2 ) }}
 D_n \cos (q|x| - \delta_e/2 ) , \nonumber \\
\Phi_{q,0} (x) & = & \sqrt{ \frac{a}{ L(| 1-b(\xi q) |^2 +| a(\xi q ) |^2 ) }}
 D_n {\rm sgn}(x) \sin (q |x| -\delta_o/2)  ,
\label{w10}
\eeqarr
where
\beqarr
\e^{\i \delta_e} & = & \frac{1+b(\xi q)}{a(\xi q)} , \nonumber \\
\e^{\i \delta_o} & = & \frac{1-b(\xi q)}{a(\xi q)} . \label{ph}
\eeqarr
We plot the phase shifts $\delta_e$ and $\delta_o$ in Fig.4(a).
 They are drawn in the region from $q=0$ to $q=\Lambda_\phi$, where we selected
$\Lambda_\phi$ as $k_F/2$. Ê
 Figure 4(b) shows the phase shifts obtained by the numerical calculation in
the SSH model.\jcite{Kurita} The agreement between the analytical result and
the numerical one is remarkablly good. The $\delta_e$ approaches $-\pi$ as $q
\rightarrow 0$, while $\delta_o$ approaches $0$ as $q \rightarrow 0$.
 The both phase shifts, $\delta_e$ and $\delta_o$ increase and approach to $0$
at the cut off $q\sim \Lambda_\phi$. These behaviors indicate that of the
electron plane waves near the edges of the electron band are pulled into the
soliton center by the attractive potential $\tilde {\zeta}(x)$.
This result is consistent with Levinson's theorem.
The theorem says that the number of localized states at the bottom of the band
reflects the phase shift.
If the numbers of localized states with even and odd parities are
$n_e$ and $n_o$, respectively, the phase shifts of the even and odd extended
states go to $ -(2n_e-1)\pi $ and $-2n_o\pi$, respectively, as $q\rightarrow
0$. \jcite{LT} The present case corresponds to $n_e=1$ and $n_o=0$.

The phase shift near the Fermi level is given by
\bequ
\e^{\i\delta} = \frac{E_\gamma}{\i +\gamma},
\label{phf}
\eequ
where $\gamma$ is the quantum number defined by eq.(\ref{w1s}). This phase
shift (\ref{phf}) is the same as that obtained in the TLM model. We have to be
careful about the definition of the wave number. At the Fermi level,
$k(=\gamma/\xi)$ is zero. Thus  $k$ is considered as the deviation of the wave
number from $k_F$. Therefore,
\bequ
k=q-k_F.
\eequ
  The phase shift $\delta$ in the valence band is plotted as the function of
$q$ in Fig.4(a). $k=0$ corresponds to $q=k_F$ and the cut off
$k=-\Lambda=-k_F/2$ corresponds to $q=k_F-\Lambda=k_F/2$, respectively.
 The phase shift $\delta$ increases as $q$ becomes larger, approaching to
$\pi/2$ as $q\rightarrow k_F$.  This behavior is also in good agreement with
the numerical result shown in Fig.4(b).
There is a discontinuity of the phase shift at $q=\Lambda_\phi$. This is
because we have used eq.(2.11) for $0 \leq q< \Lambda_\phi$ and eq.(2.13) for
$\Lambda_\phi<q\leq k_F$.

\section{Summary and Discussion}
\setcounter{equation}{0}

We have studied the effects of lattice discreteness on a soliton in the SSH
model using a modified version of the continuum (TLM) model. We solved this
model in the adiabatic limit.

When there is a soliton in the SSH model,  the arrangement of the double bonds
is changed and the ions near the soliton are shifted to the soliton center.
Then, the acoustic component of the order parameter, $\tilde{\zeta}(x)$, has
non-zero amplitude near the soliton center and it acts as a potential for the
electrons at the band edges.   Since $\zeta(x)$ acts as an attractive
potential, weakly localized states appear at the band edges.  The levels are
very close to the corresponding band edges. Because the amplitude of the
acoustic component is very small (of the order $\kappa a \Del0$), the states
are more weakly localized than the mid-gap state.
 In the TLM model, this effect (the acoustic component) is not included because
the interaction between the acoustic component $\zeta(x)$ and the electrons at
the bottom of the valence band is omitted in the continuum limit.

To improve the TLM model, we have taken into account the electronic wave
function at the bottom of the valence band and the top of the conduction band
in addition to the electron wave function at the Fermi level. We have
considered the discreteness by taking into account the terms of higher order in
$a/\xi$,
when the expansion parameter \ $a/\xi$ \ is sufficiently small.
We have studied the electronic states of electrons, using the
unperturbed Hamiltonian eqs.(\ref{r1}) and (\ref{r4}), which are composed of
the zeroth and a part of the first order terms with respect to $m_1$. The
selfconsistency between the electrons and  the order parameters have been
satisfied up to the first order of $m_1$.

The wave functions of weakly localized states are given by eq.(\ref{w9}). The
energy levels of the localized states are given by eq.(\ref{eell}).  This
result has the same form as given in the finite band version of the TLM
model.\jcite{R.S}
Concerning the scattering states,
we have obtained the dispersion relation, eqs.(\ref{Ek}) and (\ref{e2}), which
corresponds to the
scattering states in the SSH model\jcite{latex}\jcite{phase}.
\bequ
E_q=\pm\sqrt{(2t_0)^2\cos^2(q a) +\Del0^2\sin^2(q a) } .
\eequ
The wave functions are given by eqs.(\ref{w7}) and (\ref{w71}). Using these
wave functions, the phase shift analysis has been performed for the two groups
of the electronic
states, with even and odd parities respectively.
The two phase shifts coincide with each other over a wide range of the wave
number.
In the regions of small wave numbers, the two take different values which are
consistent with the fact that there are  localized states with even parity.

 Using the present model, the order parameters, eqs.(\ref{so1'}) and
(\ref{so1}), which were given by ref.\cite{solito2}, have been reproduced.
 In the expression of the electron wave functions near the band edges,
eq.(\ref{w9}), the complete set of the function $\{\phi_{q,i} (x) \}$ have been
introduced . Because of  the completeness of  $\{\phi_{q,i} (x) \}$, the
acoustic component of the order parameter, $\tilde{\zeta}(x)$, is unmodified by
the weakly localized states at the edges of the band. The order parameters are
determined by the electrons near the Fermi level.

    We have made a modified version of the TLM model and solved it
perturbationally up to the first order with respect to $a/\xi$. The present
model has a merit that it can deal with the electrons near the bottom of the
band as well as those near the Fermi level.  It has another advantage that we
can treat the interaction between the electron and the acoustic component of
the order parameter. Thus, this model is suitable for analyzing the interaction
between the acoustic phonon and the non-linear excitation. It is interesting to
study the diffusion motion of the nonlinear excitation induced by the
interaction using this model.

To obtain the exact result up to the second order of $\kappa a$, we should make
a higher order perturbation calculation.  However, since the cosine band in the
SSH model is linearized by the continuation procedure,  the contributions of
the order $\lambda\cdot  m_1$, $m_1^2$ and higher which come from the
scattering states are omitted.  Kurita {\it et al.} obtained the acoustic
component, $\tilde{\zeta}(x)$, using the finite band version of the TLM
model.\jcite{Kurita} $\tilde{\zeta}(x)$ was given by
\bequ
\tilde{\zeta}(x)  \sim
- m_1
\Biggl(
\frac{1}{2}-\frac{7}{8}\lambda
\Biggr)
\Del0\cosh^{-2}\kappa x +O(m_1^2).
\eequ
 Then, our result of the acoustic component, eq.(\ref{zeta1}), is a bit larger
than that of the SSH model. To make higher order perturbation selfconsistently,
we should construct the unperturbed Hamiltonian using the finite band version
of the TLM model.\jcite{phase}\jcite{Kurita}\jcite{R.S.P} It is a bit tedious
but important problem to explore.

{\large{\bf Acknowledgement}}

 The author would like to express his gratitude to Dr. A. Terai for his careful
reading of the manuscript and for his helpful discussions. He also express his
gratitude to Prof. Y. Wada for valuable discussions. This work was partly
supported by a Grant-in-Aid for Scientific Research from the Ministry of
Education, Science, and Culture. The numerical calculations have been performed
on the IBM RISC system 360 of Department of Physics, Faculty of Education,
Yokohama National University.

 \newpage
 \appendix

\section{Solutions of the Differential Equation (4.2)}

In this Appendix, we will consider the differential equation
\bequ
\vepsi ' \phi(x) = -\frac{\d^{2}}{\d (\kappa x)^{2}}\phi (x)-\sech^{2} (\kappa
x)
\phi(x) .
\label{eq}
\eequ
 First, the localized solution with a negative $\vepsi '$ will be  considered.
 Introducing $t (\theta)$ and $\theta$ by
\beqarr
 & & \phi_{\ell}(x) = (1-\tanh^2\kappa x )^{\frac{ \sqrt{ -\vepsi' } }{2}}
                  t( \theta )  ,\label{eq:4.2.11}   \\
 & & \theta = \frac{1+\tanh\kappa x }{2} .               \label{eq:4.2.12}
\eeqarr
%
We write eq.(\ref{eq}) in the form
\bequ
          \Bigl[
              \theta (1-\theta) \frac{\d^{2}}{\d \theta^{2}}
             +(1+\sqrt{-\vepsi '})(1-2\theta) \frac{\d}{\d\theta}
             -( 2R -\vepsi '+\sqrt{-\vepsi '})
          \Bigr] t (\theta) =0.
\label{eq:4.2.13}
\eequ
This is the hypergeometric differential equation.\jcite{Heun} With the help of
eqs.(\ref{eq:4.2.11}) and (\ref{eq:4.2.12}), the solution $\phi_{\ell}(x)$ is
given by a
hypergeometric function
\bequ
\phi_{\ell}(x)=[4 \theta (1-\theta)]^{\frac{ \sqrt{ -\vepsi '} }{2}}
          F [\sqrt{-\vepsi '}-\beta_0, \sqrt{-\vepsi '}+\beta_0+1,
	   \sqrt{-\vepsi '}+1, \theta] ,
\label{eq:4.2.14}
\eequ
where
\bequ
\beta_0=(\sqrt{5}-1)/{2}.
\label{abd}
\eequ
As we can see from eq.(\ref{eq:4.2.14}), $\phi_{\ell}(x)$ approaches zero,
as $x$ goes to $-\infty$. In order that $\phi_{\ell}(x)$ is finite in the limit
of
 \ $x \rightarrow \infty$,  the hypergeometric function $F$  should be
a polynomial. So we can put $\sqrt{-\vepsi '}-\beta_0 = -n$ with a non-negative
integer $n$. Since $\beta_0$ is about $0.6$, $n$ should be zero. Then we have
\bequ
\sqrt{-\vepsi'} = \beta_0.
\label{2b.1}
\eequ
As $\beta_0 $ is positive, we have a localized solution
\beqarr
  \phi_{\ell} (x) & = & [\theta (1-\theta)]^{\frac{ \beta_0 }{2}}
F [0, 2\beta_0+1,\beta_0+1, \theta ]
\nonumber \\
 & = & \biggl(\frac{\sech\kappa x}{2}\biggr)^{\beta_0} .
\label{eq:sol2}
\eeqarr

The eigen value $\varepsilon'$ is given by
\bequ
\varepsilon'   =  -\beta_0^{2}/2 .
\label{ep2}
\eequ
 To obtain the solutions of scattering states with a positive $\vepsi '$, we
replace \ $\sqrt{-\varepsilon'}$
 \ in eq.(\ref{eq:4.2.14})
 \ by \ $\i \gamma$, that is $\vepsi'=\gamm^2$, to get
\bequ
\phi_{\gamma}(x)    =
 [\theta (1-\theta)]^{\frac{ \i\gamma}{2}}
 F [-\beta_0+\i\gamma, \  \beta_0+1+\i\gamma, \
 1+\i\gamma ; \ \theta ] .
 \label{eq:4.2.16}
\eequ
The hypergeometric function has the  following relations
\beqarr
F[\al, \beta, \gamma ; \theta] & = &
      (1-\theta)^{\gamma-\al-\beta}
     F [\gamma-\al, \gamma-\beta, \gamma ; \theta ],
   \label{eq:4.2.18}  \\
 & & \nonumber \\
 F[\al, \beta, \gamma ; \theta] & = &
     \frac{ \Gamma(\al+\beta-\gamma) \Gamma(\gamma) }
          { \Gamma(   \al   ) \Gamma(  \beta  )}
      (1-\theta)^{\gamma-\al-\beta}
     F [\gamma-\al, \gamma-\beta, \gamma-\al-\beta+1 ; 1-\theta ]
   \nonumber    \\
     &  & +   \frac{ \Gamma(\gamma) \Gamma(\gamma-\al-\beta) }
                  { \Gamma( \gamma-\al ) \Gamma( \gamma-\beta )}
        F [\al, \beta, \al+\beta-\gamma+1 ; 1-\theta ] .
      \label{eq:4.2.19}
\eeqarr
We write $\phi_{\gamm}$ as a function of \ $1-\theta$, \ using
eqs.(\ref{eq:4.2.18})
and (\ref{eq:4.2.19}) and the relation
$ \e^{-2z}=\frac{1-\theta}{\theta}$ to get
\beqarr
\phi_{\gamm} (x) & = &   \e^{\i \gamm x} F [1+\beta_0, -\beta_0, 1+\i\gamm ;
\theta ]
             \nonumber \\
    & = & b(\gamm) \phi_{\gamm}(-x)+a(\gamm)\phi_{-\gamm}(-x),
\label{eq:4.2.20}
\eeqarr
where
\beqarr
a(\gamm) & = & \frac{ \Gam(1+\i\gamm)\Gam( \i\gamm) }
                   { \Gam(1+\beta_0 +\i\gamm)\Gam(-\beta_0 +\i\gamm) },
	     \nonumber \\
b(\gamm) & = & \frac{ \Gam(1 +\i\gamm)\Gam( - \i\gamm) }
                   { \Gam(1+\beta_0)\Gam(-\beta_0) } .
	    \label{eq:4.2.21}
\eeqarr
We rewrite $\phi_\gamma (x)$ by $q=\gamma/\xi$,
\bequ
\phi_q (x) =\phi_\gamma (x) = \e^{\i q x} F[1+\beta_0,-\beta_0,1+\i\xi q;
\theta],
\label{eq:4.2.20s}
\eequ
$q$ is the wave number of the electron at the band edges.
{}From eqs.(\ref{eq:4.2.20}) and (\ref{eq:4.2.20s}), we can see that the
function $\phi_{\gamm}$  has the asymptotic forms
\bequ
    \phi_{\gamma}(x) = \left \{ \begin{array}{ll}
     \e^{ \i q x}               &  {\rm as}\  x \rightarrow -\infty , \\
      a(\xi q) e ^{ \i q x} + b(\xi q) e ^{  -\i \xi q}
      &   {\rm as}\  x \rightarrow  \infty .
     \end{array}     \right.
\label{eq:4.2.22}
\eequ

\section{Calculation of Normalization Factor}

Consider an hermitian operator $L_{0}$ and its eigen function $u_{\gamm}(x)$

\beqarr
 & & L_{0}         =   -\frac{{\rm d}^{2}}{{\rm d}z^{2}}+U(z) ,\nonumber \\
 & & L_{0} u_{\gamm} = E(\gamm)u_{\gamm} ,
\label{A.1}
\eeqarr
where the  eigenvalue $E(\gamm)$ is related to the quantum number $\gamm$ by
$E(\gamm)=\gamm ^{2}$.
We differntiate eq.(\ref{A.1}) by $\gamma$ and multiply by $u_{-\gamm}^*$ from
the left to obtain
\bequ
 u_{\gamm}^*(L_{0} -\gamm^{2})u'_{\gamm}=2\gamm u_{\gamm}^*u_{\gamm} .
\label{A.3}
\eequ
The hermitian conjugate expression of eq.(\ref{A.1}) gives

\bequ
(L_{0}u_{\gamm}^{*}-\gamm^2  u_{\gamm}^{*})u'_{\gamm} =  0 .
\label{A.2}
\eequ
The difference between the two is integrated to give
\bequ
\Biggl[\bigl(\frac{{\rm d}}{{\rm d}z}u^{*}_{\gamm}\bigr)u'_{\gamm}
 - u_{\gamm}^{*} \bigl(\frac{{\rm d}}{{\rm d}z}u'_{\gamm}\bigr) \Biggr]^{a}_{b}
 = 2\gamma \int^{a}_{b} u^{*}_{\gamm}u_{\gamm} {\rm d}z .
 \label{A.4}
 \eequ
  Suppose we put
\bequ
u_\gamma = \phi_{\gamma}(x)\pm \phi_{\gamm}(-x),
\eequ
which give $\phi_{q,i}(x)$.
With the help of eq.(\ref{eq:4.2.22}), we get
 \beqarr
 \int^\frac{L}{2}_\frac{-L}{2} |\phi_{k,i}(x)|^2 \d x &  = &
\int^\frac{L}{2}_\frac{-L}{2} |u_{\gamma}(x)|^2 \d x
\nonumber \\
& = & (|1\pm b(\gamm)|^2+ |a(\gamm)|^2 )L +O(1)
 \label{A.9}
 \eeqarr
%
 The wave functions of the localized states eq.(\ref{eq:4.2.14}) can be
obtained
 from $ u_{\gamm}(x)$, if we perform an analytic continuation
 \bequ
 \gamm \rightarrow \gam1 \equiv i(\al-n),
 \nonumber
 \label{A.6}
 \eequ
 where $n$ is a non-negative integer. Equation (\ref{A.4}) becomes
\bequ
\Biggl[ \bigl(\frac{{\rm d}}{{\rm d}z}u_{\gam1}\bigr)u'_{\gam1}
 - u_{\gam1} \bigl(\frac{{\rm d}}{{\rm d}z}u'_{\gam1}\bigr) \Biggr]^{a}_{b}
 = 2\gam1 \int^{a}_{b} (u_{\gam1})^2 {\rm d}z .
 \label{A.7}
 \eequ
 Since eq.(\ref{eq:4.2.21}) gives
$$
\left. \begin{array}{l}
     a(\gam1)=0, \\
     b(\gam1)={\rm i}(-1)^{n},  \ \ \ \ \mbox{for }\gam1=\i\beta_0
      \end{array}  \right.
$$
substitution of eq.(\ref{eq:4.2.22}) with $\gamma=\gam1$ into eq.(\ref{A.7})
gives
 \beqarr
 N^{-2} & = & \int^{L/2}_{-L/2}(\phi_{ 1,\gam1 }(x))^{2}{\rm d}x   \nonumber \\
        & = & {\rm i} a'(\gam1)b(\gam1)/\kappa       \nonumber \\
        & = & \frac{n!}{\kappa}
          \frac{\Gamma(\beta_0 +1-n)\Gamma(\beta_0 -n)}{\Gamma(1+2\beta_0 -n)}
{}.
 \label{A.8}
 \eeqarr
In particular, for $ \phi_{\ell} $
\beqarr
 \int^{L/2}_{-L/2} \cosh(\kappa x)^{-2\beta_0} dx
        & = & \frac{2^{2\beta_0} n!}{\kappa}
          \frac{\Gamma(\beta_0+1)\Gamma(\beta_0)}{\Gamma(1+2\beta_0)} .
 \label{A.11}
\eeqarr
%
\newpage


\begin{figcaption}

\item Levels of localized states of a soliton.

Two states, B and C, are close to the band edges and weakly localized.
The energy of the these two states are given by eq.(\ref{eell}). The state A is
a mid-gap state.

\item  Electron density distributions of the bound state wave functions for the
coupling constant $\lambda=0.2$.

The ordinate is scaled by $1/\xi$. The abscissa is $x/\xi$.  $|\Phi_\ell|^2$ is
the electron density at the bottom of the valence band.
$|\Phi_M|^2$ is the electron density in the mid-gap localized state


\item Schematic electronic band structure of a soliton.

The dispersion relation at the Fermi level, eq.(\ref{Ek}), is plotted as the
function of $k=q-k_F$, which is identical with that of the TLM model. The
dipsersion relation at the band edges, eq.(\ref{e2}),  have a parabolic form,
which is plotted as the function of $q$. We selected the cut off as
$\Lambda=\Lambda_\phi=k_F/2$.

\item  Phase shifts of the wave functions as a functions of $q$. The
dimensionless coupling constant $\lambda$ is selected as $0.2$. The ordinate is
scaled by $1/\pi$. The abscissa is  $q/k_F$.

a)  The phase shifts $\delta_{e}$ and $\delta_{o}$ given by eq.(\ref{ph}).
 $\delta$ is the phase shift of electrons near the fermi level.  We selected
the cut off as $\Lambda=\Lambda_\phi=k_F/2$.
b) The phase shift of electrons in the SSH model obtained in ref.\cite{Kurita}.

\end{figcaption}

\end{document}